\documentclass{article}

\usepackage{PRIMEarxiv}

\usepackage[utf8]{inputenc} 
\usepackage[T1]{fontenc}    
\usepackage{hyperref}       
\usepackage{url}            
\usepackage{booktabs}       
\usepackage{amsfonts}       
\usepackage{nicefrac}       
\usepackage{microtype}      
\usepackage{lipsum}
\usepackage{fancyhdr}       
\usepackage{graphicx}       
\graphicspath{{media/}}     
\usepackage[algo2e,ruled,vlined]{algorithm2e}
\usepackage{cite}
\usepackage{amsmath,amssymb,amsfonts}
\usepackage{graphicx}
\usepackage{textcomp}
\usepackage{xcolor}
\usepackage{subfig}
\usepackage{amssymb}
\usepackage{multirow}
\usepackage{multicol}
\title{Speed-enhanced Subdomain Adaptation Regression \\
for Long-term Stable Neural Decoding \\
in Brain-computer Interfaces
}

\author{
  Jiyu Wei, Dazhong Rong, Xinyun Zhu, Qinming He, Yueming Wang  \\
  College of Computer Science and Technology \\
  Zhejiang University \\
  Hangzhou, China\\
  \texttt{\{weijiyu, rdz98, zhuxinyun, hqm, ymingwang\}@zju.edu.cn} \\
}

\begin{document}
\maketitle

\begin{abstract}
Brain-computer interfaces (BCIs) offer a means to convert neural signals into control signals, providing a potential restoration of movement for people with paralysis. Despite their promise, BCIs face a significant challenge in maintaining decoding accuracy over time due to neural nonstationarities.  However, the decoding accuracy of BCI drops severely across days due to the neural data drift. While current recalibration techniques address this issue to a degree, they often fail to leverage the limited labeled data, to consider the signal correlation between two days, or to perform conditional alignment in regression tasks. This paper introduces a novel approach to enhance recalibration performance. We begin with preliminary experiments that reveal the temporal patterns of neural signal changes and identify three critical elements for effective recalibration: global alignment, conditional speed alignment, and feature-label consistency. Building on these insights, we propose the Speed-enhanced Subdomain Adaptation Regression (SSAR) framework, integrating semi-supervised learning with domain adaptation techniques in regression neural decoding.  SSAR employs Speed-enhanced Subdomain Alignment (SeSA) for global and speed conditional alignment of similarly labeled data, with Contrastive Consistency Constraint (CCC) to enhance the alignment of SeSA by reinforcing feature-label consistency through contrastive learning. Our comprehensive set of experiments, both qualitative and quantitative, substantiate the superior recalibration performance and robustness of SSAR.
\end{abstract}

\begin{figure*}
        \centering
        \includegraphics[width=1\linewidth]{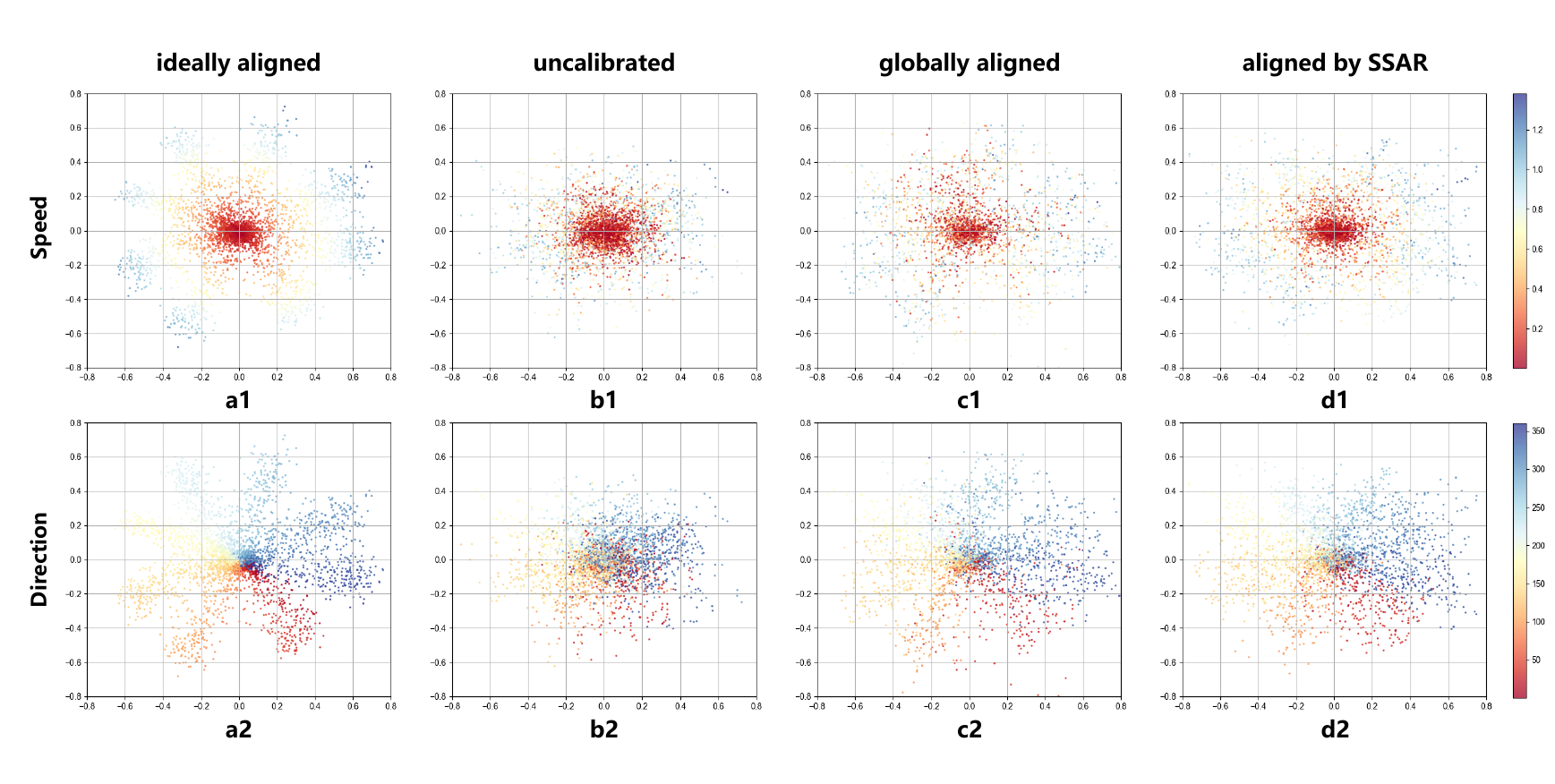}
	\caption{Figure1: Feature distributions with different alignments colored by speed or direction of ground-true velocity. \textbf{a)} An ideal feature distribution exhibits
a radial shape and preserves label-feature consistency. \textbf{b)} Significant deviations occur in uncalibrated feature distribution. \textbf{c)} Previous methods perform global alignment mitigating the deviations to some extent but fail in conditional speed alignment. \textbf{d)} Our SSAR tackles the weakness of previous methods and performs a near-ideal alignment.}
        \label{fig1}
\end{figure*}

\keywords{recalibration \and neural decoding \and brain-computer interfaces \and semi-supervised domain adaptation regression}

\section{Introduction}~\label{sec:intro}
Brain-computer interfaces (BCIs) facilitate the communication between neural processes and external devices by decoding neural activity into intentional commands, which have demonstrated great potential in restoration, rehabilitation, and improvement for individuals with disabilities\cite{bci1,bci2,bci3,bci4,restore1}. 
Numerous neural decoding methods, such as PV\cite{PD}, Kalman Filter\cite{kalman}, KRSRL\cite{KRSRL}, DyEnsemble\cite{DyEnsemble} and EvoEnsemble\cite{EvoDyEnsemble}, have been proposed, achieving impressive decoding accuracy. 
Concurrently, the evolution of neuron population recording techniques has expanded the capacity for complex interactions with the external environment, thereby broadening the applicability of BCIs in real-world contexts\cite{neuronreocding1,neuronreocding2,neuronreocding3,neuronreocding4}.
Despite these advancements, BCIs encounter a significant challenge: the variability of neural signals over time due to factors like electrode drift, neuronal turnover, and synaptic plasticity\cite{brain1,brain2,Brainplasticity}. 
These changes undermine the stability of decoding performance, posing a substantial barrier to the long-term utility of BCIs\cite{brain2,long}.

To address this issue, various recalibration methods have been proposed. A common approach is supervised daily recalibration, which retrains the decoder during use\cite{daily,daily2}. 
However, acquiring large amounts of labeled data in real-world applications is often difficult and expensive. 
Although some semi-supervised learning (SSL) methods\cite{semibci,meanteacher} have been adopted for BCIs recalibration, they often overlook the utilization of data from the preceding days, suggesting opportunities for further refinement.
Recently, unsupervised domain adaptation (UDA) has achieved notable success in computer vision\cite{udacv1} and natural language processing\cite{udanlp}, prompting its adoption in BCIs recalibration through methods like CCA\cite{long}, ADAN\cite{adan}, CycleGAN\cite{cyclegan}, WDGRL\cite{WDGRL} and DA-DCF\cite{DA-DCF}.
Because these methods do not make use of the few labeled data in real-world recalibration scenarios, they often struggle to achieve optimal performance.
A promising approach to BCIs recalibration involves applying semi-supervised domain adaptation (SSDA), which has recently received considerable attention for its ability to enhance performance at a low-cost significantly\cite{ssda1,ssda2,ssda3}.
However, most existing SSDA methods are designed for classification tasks and are not suitable for regression-based neural decoding tasks.
Currently, no SSDA-based recalibration method has been proposed for BCIs applications.

Moreover, to explore the patterns of neural signal changes over days, we conduct a series of preliminary experiments from the views of speed and direction of ground-true velocity (\textit{i.e.,} the labels).
Our observations reveal that an ideal feature distribution exhibits a radial shape and preserves consistency between labels and features.
When extracting features of day-$k$ data using a decoder trained on day-$0$ data, we notice significant deviations in speed decoding, while direction decoding remains relatively stable.
This means the features from two days with similar directions are still close, and hence the small deviations in direction can be corrected well by previous UDA methods with global alignment, mitigating performance degradation to some extent.
But due to the drastic change in the patterns about speed\cite{ps,EvoDyEnsemble}, the features from two days with similar speeds may be far apart, which is beyond the correction capabilities of the global alignment.
Hence, these methods failed in conditional speed alignment, which has been suggested to be beneficial by recent studies\cite{dsan,cmmd,smmd}, resulting in limited recalibration performance.
Besides, existing conditional alignment approaches are usually designed for classification tasks, which are not suitable for our regression-based neural decoding tasks. 

Based on these findings, we propose three key factors for accurate recalibration: global alignment, conditional speed alignment, and feature-label consistency. Building on these factors, we introduce the Speed-enhanced Subdomain Adaptation Regression (SSAR) framework, which is the first to consider domain adaptation and semi-supervised learning for regression-based recalibration. Specifically, we propose Speed-enhanced Subdomain Alignment (SeSA) to perform global and conditional speed alignment by aligning data with similar labels. Besides, Contrastive Consistency Constraint (CCC) is proposed to further enhance the alignment of SeSA by enforcing consistency between features and labels in a contrastive learning manner.


\begin{figure}[t]
\centerline{\includegraphics[width=0.8\linewidth]{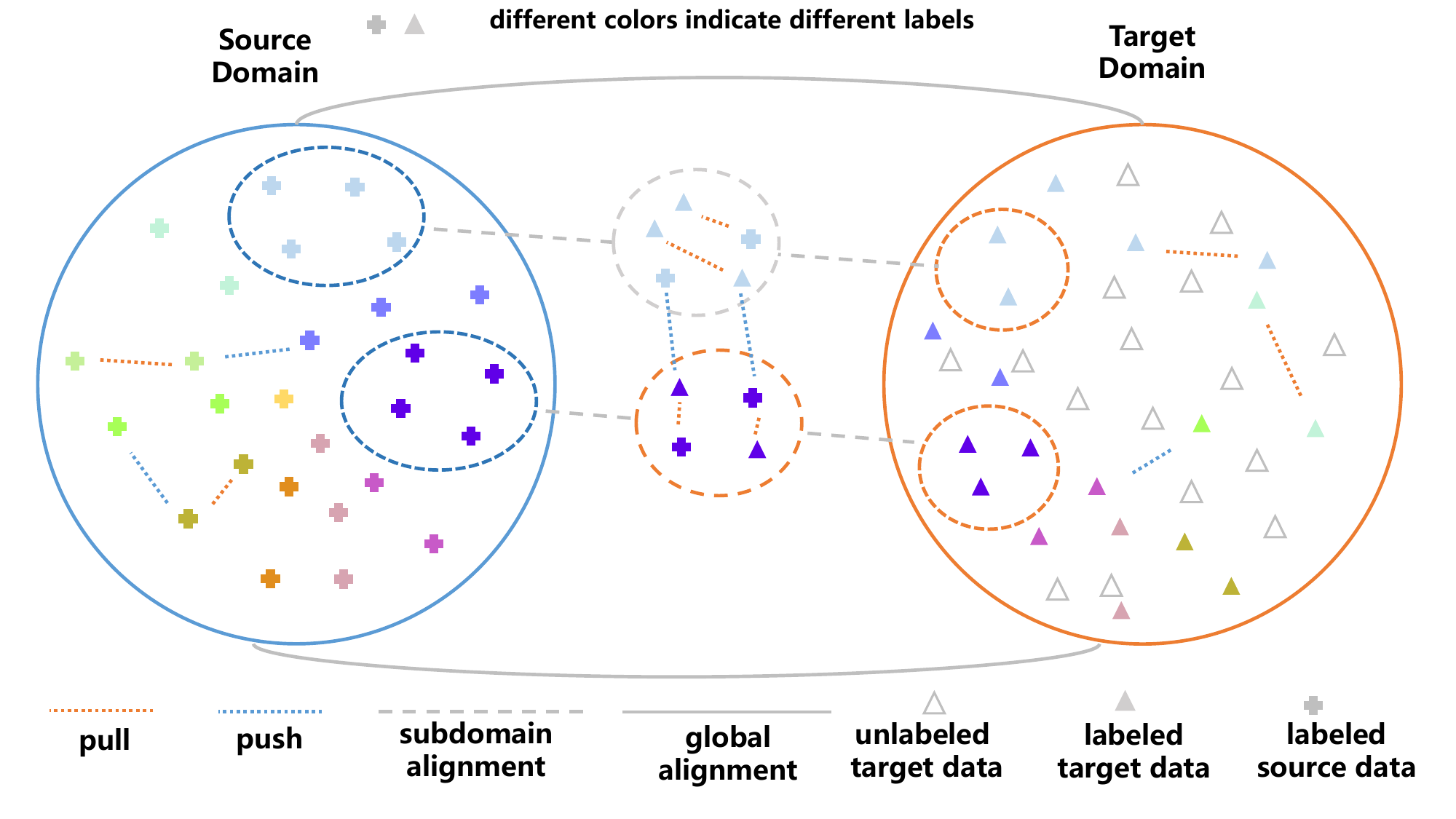}}
\caption{SeSA and CCC.}
\vspace{-15px}
\label{fig2}
\end{figure}

Our contributions can be summarized as follows:
\begin{itemize}
\item We conduct preliminary experiments to explore the patterns of neural signal changes over days and find the lack of conditional speed alignment limiting the performance of previous UDA methods for recalibration. Then, we propose three key factors to improve recalibration performance: global alignment, conditional speed alignment, and feature-label consistency.
\item Based on the three keys, we propose a novel recalibration method named Speed-enhanced Subdomain Adaptation Regression (SSAR). To the best of our knowledge, this is the first approach to consider semi-supervised learning and subdomain adaptation for regression-based BCIs recalibration, balancing performance and cost.
\item We conduct extensive qualitative and quantitative experiments, demonstrating that our method achieves state-of-the-art recalibration performance.
\end{itemize}

\section{Related work}
The BCIs recalibration aims to maintain the decoding performance on day-k of the decoder trained on day-0.
From the aspect of the utilization of day-k data, existing studies can be divided into two categories: conventional recalibration and domain adaptation based recalibration.

\subsection{Conventional Recalibration Methods}
For maintaining decoding accuracy, the conventional recalibration methods directly utilize the labeled day-k data.
Specifically, these methods usually periodically collect labeled data on day-k and retrain the decoder on the collected data\cite{daily,daily2}.
The effectiveness of these methods is highly dependent on a large amount of the labeled day-k data.
However, collecting these data is often difficult and time-consuming, which limits the practicability of these methods.

Besides, some semi-supervised learning methods are used to recalibrate BCIs as well.
These methods train the decoder on the unlabeled day-k data in addition to the labeled day-k data by semi-supervised learning, which requires less labeled data hence the better practicability\cite{semibci,meanteacher}.
All these conventional recalibration methods do not take into account the relationship between day-0 and day-k, which limits their recalibration performance.

\subsection{Domain Adaptation Based Recalibration Methods}
With the success of transfer learning in computer vision\cite{udacv1}, natural language processing\cite{udanlp}, etc., some recent works have begun to recalibrate BCIs by domain adaptation.
Different from the conventional recalibration methods, these methods assume all day-k data are unlabeled, which means they do not make use of any labels on day-k.
These methods consider the day-0 data as the source domain and the day-k data as the target domain.
Specifically, CCA\cite{long} aims to linearly map the features of the source domain to the target domain.
However, CCA requires one-to-one pairs of samples between the two domains, which restricts its application to real-world scenarios.

Some other domain adaptation based recalibration methods are data-driven and work in a similar way to style transfer.
ADAN\cite{adan} achieves the transformation between the source and target domains in an adversarial learning manner by minimizing the VAE reconstruction loss of the two domains.
CycleGAN\cite{cyclegan}, another adversarial learning based algorithm, achieves better performance by transforming the source domain into the target domain and the target domain into the source domain through cycles.
DA-DCF\cite{DA-DCF} and WDGRL\cite{WDGRL} align the source and target domains from the perspective of features.
DA-DCF aligns the features of the two domains by confusing the discriminator to make the features of the two domains more indistinguishable. 
WDGRL adopts several common methods to measure the feature distance between the two domains and performs the alignment by minimizing the feature distance.
Although the above domain adaptation based recalibration methods achieve good results by globally aligning the source and target domains, they ignore the conditional alignment which results in their failure in speed decoding.
Recent studies have proved that conditional alignment is helpful for accurate recalibration\cite{dsan,cmmd,smmd}.
However, these methods are designed for classification-based tasks.

Our proposed method can be regarded as semi-supervised domain adaptation regression.
Unlike existing methods, our SSAR takes use of both the unlabeled data and few labeled data in target domain.
Besides, SSAR performs conditional speed alignment for regression-based BCIs recalibration.
Note that our method is closer to real-world scenarios, which can significantly improve the recalibration performance with few easily accessible labeled target data\cite{ssda1,ssda2,ssda3}.

\section{PRELIMINARY} \label{sec:pre}

\subsection{Problem Definition}
Neural decoding aims to translate the neural activities into control signals. 
Specifically, let $\mathcal{D}_k$ denote the data set collected on day $k$.
For each $(x_k, y_k)\in\mathcal{D}_k$, $x_k$ and $y_k$ denote the vector of neural activities and the vector of control signals, respectively.
The predicted vector of control signals is denoted as $\hat{y}_k=\Phi_k(\Psi_k(x_k))$, where $\Psi_k$ and $\Phi_k$ denote the feature extractor and the regressor trained on $\mathcal{D}_k$, respectively.


Considering the realities referred to in Section~\ref{sec:intro}, we regard our proposed recalibration method as a semi-supervised domain adaptation regression.
Specifically, we treat $\mathcal{D}_i$ and $\mathcal{D}_j$ as the source domain $\mathcal{D}_s$ and the target domain $\mathcal{D}_t$, where $i<j$.
In recalibration, we are given not only the source data $\mathcal{D}_s=\{(x_s^i,y_s^i)\}_{i=1}^{N_s}$, but also the unlabeled target data $\mathcal{D}^u_t=\{(x_u^i)\}_{i=1}^{N_t^u}$ and few easily accessible labeled target data $\mathcal{D}^l_t=\{(x_l^i,y_l^i)\}_{i=1}^{N_t^l}$.
Note that $N_s$ denotes the number of samples in the source domain.
$N_t^u$ and $N_t^l$ denote the number of unlabeled samples and labeled samples provided for recalibration in the target domain, where $N_t^u \gg N_t^l$.


\begin{figure*}[htb!]
\centerline{\includegraphics[width=0.95\linewidth]{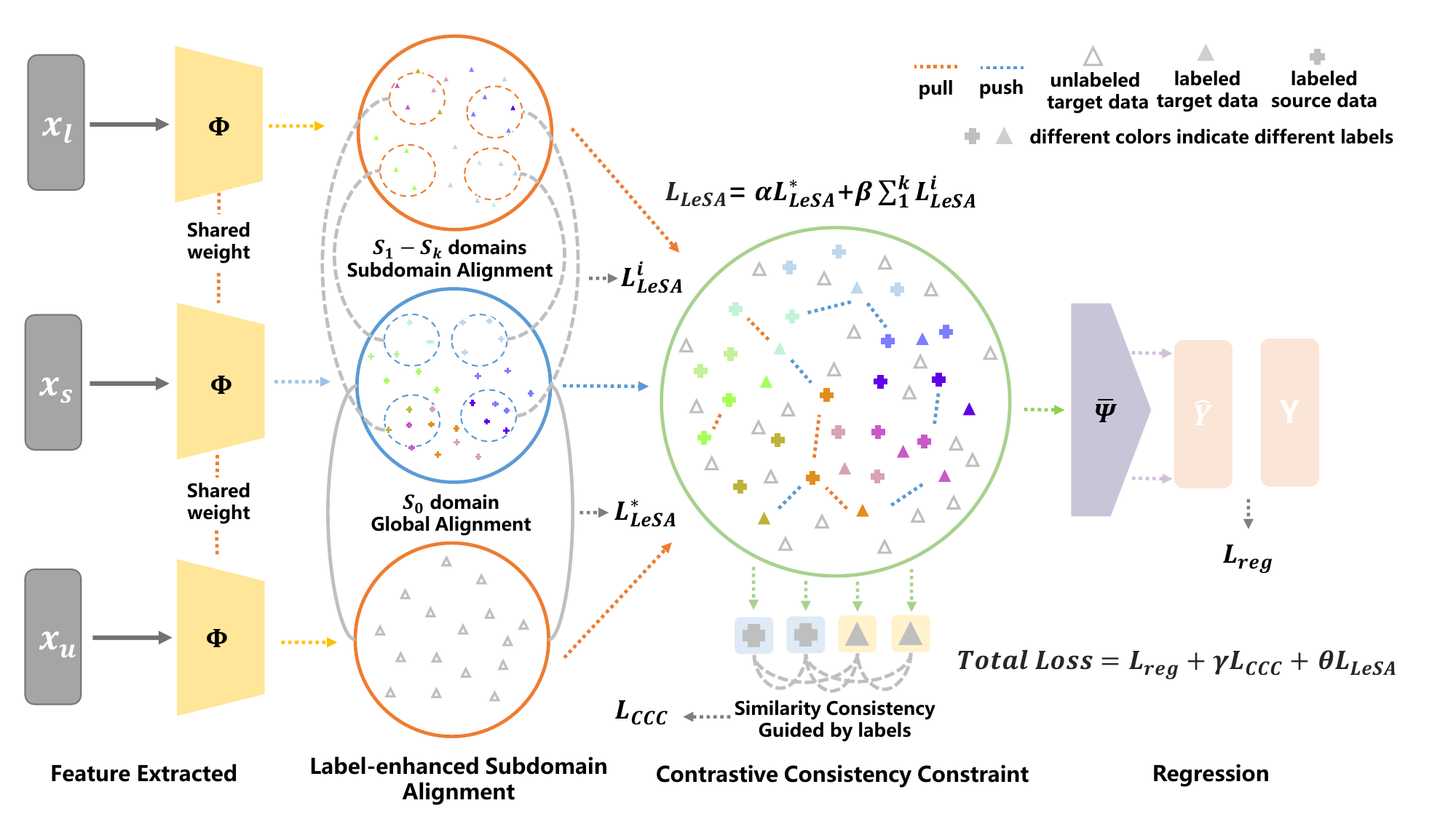}}
\caption{The framework of Speed-enhanced Subdomain Adaptation Regression. }
\vspace{-10px}
\label{fig3}
\end{figure*}

\subsection{Observed Potential Patterns across Days}
\subsubsection{Settings} For better recalibration performance, we try to observe the potential patterns by diving into the feature distributions across days.
When recalibrating the decoder from $\mathcal{D}_s$ to $\mathcal{D}_t$, we first obtain the well-trained feature extractor $\Psi_s$ and regressor $\Phi_s$ on $\mathcal{D}_s$.
Afterwards, we \textbf{fix the regressor as $\Phi_s$} and train the feature extractor $\Psi_t$ on $\mathcal{D}_t$.
Finally we reduce the dimensionality of $\{\Psi_s(x_t^i)\}_{i=1}^{N_t}$ and $\{\Psi_t(x_t^i)\}_{i=1}^{N_t}$ by Principal Components Analysis (PCA) for visualization.
The results of $\{\Psi_s(x_t^i)\}_{i=1}^{N_t}$ are plotted on Fig.~\ref{fig1}(b1,b2).
And the results of $\{\Psi_t(x_t^i)\}_{i=1}^{N_t}$ are plotted on Fig.~\ref{fig1}(a1,a2).
Note that in Fig.~\ref{fig1}, the point color of $x_t^i$ indicates the direction angle and speed of $y_t^i$, respectively.

\subsubsection{Potential Patterns across Days}
For each $x_t^i$, $\Psi_t(x_t^i)$ is its ideal extracted feature, and $\Psi_s(x_t^i)$ is its uncalibrated feature which is naively extracted by $\Psi_s$.
From the observation of Fig.~\ref{fig1}(a1,a2), the ideal feature distribution exhibits a star-like radial shape, and the relative relationships among the labels are well preserved (\textit{i.e.,} the similarity of the extracted features and the similarity of the labels are generally positively correlated), which indicates the consistency between extracted features and labels.
By contrast, in Fig.~\ref{fig1}(b1,b2) the extracted feature distribution collapses into a cluster.
The points with different directions approximately maintain the distinguishability, but the points with different speeds are mixed together and lose the above consistency.
In conclusion, the global distribution collapse and the feature-label inconsistency finally result in performance degradation across days.

Existing domain adaption methods are designed to roughly align the feature distributions across days, which can mitigate the performance degradation to some extent\cite{WDGRL,adan,cyclegan,DA-DCF}.
For a deeper understand of these methods, we adopt MMD\cite{mmd1}, one of the most effective and widely used domain adaptation methods, to align $\Psi_s(x_t^i)$ and $\Psi_s(x_s^i)$.
Same as before, we visualize the distribution of $\Psi_s(x_t^i)$ with MMD in Fig.~\ref{fig1}(c1,c2).
We can observe that the collapsed points are separated and the global feature distributions of different days are aligned roughly with the directional feature-label consistency.
However, the points with different speeds are still mixed together especially in low-speed areas, indicating the absence of conditional speed alignment due to the lack of labels.

From these observations, we conclude three possible keys for accurate recalibration: \textbf{i)} global alignment; \textbf{ii)} conditional speed alignment; \textbf{iii)} feature-label consistency.

\section{Method}
Considering the limitation of unsupervised domain adaptation and semi-supervised learning in more realistic BCIs applications\cite{semibci,meanteacher,adan,cyclegan,DA-DCF,WDGRL}, we propose 
Speed-enhanced Subdomain Adaptation Regression (SSAR) for recalibration, as shown in Fig.~\ref{fig3}.
Based on the three keys for improving recalibration performance discussed in Section~\ref{sec:pre}, we propose two components, named Speed-enhanced Subdomain Alignment (SeSA) and Contrastive Consistency Constraint (CCC).

\subsection{Speed-enhanced Subdomain Alignment}
As observed in Fig.~\ref{fig1}(a,c), after the global alignment of existing domain adaptation methods, the direction distribution of the aligned features is similar to that of the ideal features, but the speed distribution of the aligned features is distorted which is quite different from that of the ideal features.
We argue this distortion in the speed distribution is the major reason for the degradation of recalibration performance.
Therefore, considering in real-world recalibration scenarios few labeled data in the target domain are usually easily accessible, we propose Speed-enhanced Subdomain Alignment (SeSA) which utilizes them to further align the speed distribution conditionally based on global alignment.

In SeSA, for conditional speed alignment, we first divide the source data $\mathcal{D}_s$ and the labeled target data $\mathcal{D}^l_t$ into different subdomains.
Specifically, we define $C$ subdomains, denoted by $\mathcal{M}_1, \mathcal{M}_2, \dots, \mathcal{M}_C$ respectively.Then we compute the maximum speed, the minimum speed, and the speed interval as $v_{\text{max}}=\max_{(x,y)\in\mathcal{D}_s\cup\mathcal{D}^l_t} \|y\|_2$, $v_{\text{min}}=\min_{(x,y)\in\mathcal{D}_s\cup\mathcal{D}^l_t} \|y\|_2$ and $\Delta=\frac{v_{\text{max}}-v_{\text{min}}}{C}$, respectively.
For each $(x,y)\in\mathcal{D}_s\cup\mathcal{D}^l_t$, $x\in\mathcal{M}_i$ if $\frac{\|y\|_2-v_{\text{min}}}{\Delta}\in[i-1,i]$, where $i\in\{1,2,\dots,C\}$.
Specially, for convenience, we define an extra subdomain $\mathcal{M}_*=\{x_s|(x_s,y_s)\in\mathcal{D}_s\}\cup\mathcal{D}^u_t$ for global alignment.
Maximum Mean Discrepancy (MMD)~\cite{mmd1} is one of the most widely used measurements to evaluate the discrepancy between two distributions.
Different from existing MMD-based methods~\cite{mmd2,dsan,cmmd,smmd} simply adopting MMD to align the distributions globally, our proposed SeSA adopts MMD to weightedly align the distributions between the source data and the target data within each predefined subdomain.
For each subdomain $\mathcal{M}_i$ ($i\in\{1,2,\dots,C,*\}$), we define the source subdomain as $\mathcal{M}_i^s=\{x|(x,y)\in\mathcal{D}_s\land x\in\mathcal{M}_i\}$ and the target subdomain as $M_i^t=\{x|(x,y)\in\mathcal{D}_t\land x\in\mathcal{M}_i\}$.
The corresponding subdomain loss is defined as follows:
\begin{equation}
    \mathcal{L}_\text{SeSA}^i=\left\| \frac{1}{|\mathcal{M}_i^s|} \sum_{x_s\in\mathcal{M}_i^s} \Psi(x_s) - \frac{1}{|\mathcal{M}_i^t|} \sum_{x_t\in\mathcal{M}_i^t}  \Psi(x_t) \right\|^2_{\mathcal{H}},
    \label{eq1}
\end{equation}
where $\mathcal{H}$ is the reproducing kernel Hilbert space (RKHS) endowed with a characteristic kernel $\mathbb{K}$.
For clarity, we can reformulate the above equation as:
\begin{equation}
    \begin{aligned}
        \mathcal{L}_\text{SeSA}^i & =\Bigg\| \frac{1}{|\mathcal{M}_i^s|^2} \sum_{x_s^j\in\mathcal{M}_i^s} \sum_{x_s^k\in\mathcal{M}_i^s} \mathbb{K}(x_s^j, x_s^k) +\\
        &\quad\quad \frac{1}{|\mathcal{M}_i^t|^2} \sum_{x_t^j\in\mathcal{M}_i^t} \sum_{x_t^k\in\mathcal{M}_i^t} \mathbb{K}(x_t^j, x_t^k) -\\
        &\quad\quad \frac{2}{|\mathcal{M}_i^s||\mathcal{M}_i^t|} \sum_{x_s^j\in\mathcal{M}_i^s} \sum_{x_t^k\in\mathcal{M}_i^t} \mathbb{K}(x_s^j, x_t^k)
        \Bigg\|.
    \end{aligned}
        \label{eq2}
\end{equation}

Finally, we define the complete loss of SeSA in a unified form as follows:
\begin{equation}
\mathcal{L}_\text{SeSA}=\alpha\cdot\mathcal{L}_\text{SeSA}^*+\beta\cdot\frac{1}{C}\cdot\sum_{i=1}^C \mathcal{L}_\text{SeSA}^i.
\end{equation}
Note that both $\mathcal{L}_\text{SeSA}^*$ and $\mathcal{L}_\text{SeSA}^i$ in the equation aim to minimize the feature distribution discrepancy in the corresponding subdomains.
Factually, $\mathcal{L}_\text{SeSA}^*$ leads to the aforementioned global alignment, and these $\mathcal{L}_\text{SeSA}^i$ result in the aforementioned conditional speed alignment.
The hyperparameters $\alpha$ and $\beta$ are the weights used to balance global alignment and conditional speed alignment.

\subsection{Contrastive Consistency Constraint}
In recent studies, contrastive learning has demonstrated its high effectiveness under single domain settings on various fields~\cite{KRSRL,ssda2}.
The core idea of contrastive learning is intuitive that samples with similar labels should exhibit similar features.
Some existing works have extended contrastive learning under multi-domain settings and have achieved encouraging results\cite{ssda2}.
However, these works only take the classification tasks into account, and can not be directly generalized to regression-based neural decoding.

From the observation of Fig.~\ref{fig1}(a), we can conclude an essential characteristic of ideal feature distributions: feature-label consistency.
Constraining feature-label consistency can further enhance the alignment of SeSA because it preserves the relative relationships among samples.
To obtain better feature-label consistency, we propose Contrastive Consistency Constraint (CCC).
In CCC we combine the source data and the labeled target data as $\mathcal{D}^l=\mathcal{D}_s\cup\mathcal{D}_t^l$, and constrain the feature-label consistency in $\mathcal{D}^l$ by a contrastive manner.
Specifically, we first define the feature similarity between $x_1$ and $x_2$, and the label similarity between $y_1$ and $y_2$.
For simplicity, we adopt the local-response Gaussian kernel\cite{adan} for both similarities, as follows:
\begin{equation}
\begin{aligned}
&\Omega(x_1, x_2)=\mathrm{e}^{-\frac{\left\|x_1-x_2\right\|_2^2}{2}},\\
&\Omega(y_1, y_2)=\mathrm{e}^{-\frac{\left\|y_1-y_2\right\|_2^2}{2}}.
\end{aligned}
\label{eq3}
\end{equation}
The aforementioned feature-label consistency indicates that, for any two samples if their labels are similar their extracted features should be similar.
Hence, we define the following loss to contrastively constrain the feature-label consistency:
\begin{equation}
\mathcal{L}_{\text{CCC}}=\sum_{(x_1,y_1)\in\mathcal{D}^l} \sum_{(x_2,y_2)\in\mathcal{D}^l} \big(\Omega(x_1,x_2)-\Omega(y_1,y_2)\big)^2.
\label{eq4}
\end{equation}

\subsection{Overall Framework}
In the previous two subsections, the Speed-enhanced Subdomain Alignment (SeSA) addressed the challenges of global alignment and conditional speed alignment.
The Contrastive Consistency Constraint (CCC) solves the problem of the feature-label consistency which enhances the alignment of SeSA.
Combining these two components, in this subsection we propose our overall framework named Speed-enhanced Subdomain Adaptation Regression (SSAR) for BCIs recalibration. 

To obtain the well-adapted feature extractor $\Phi_t$ and regressor $\Psi_t$ on the target domain, we first randomly initialized the learnable parameters of $\Phi_t$ and $\Psi_t$.
Then we optimize these parameters by minimizing the complete recalibration loss as follows:
\begin{equation}
\mathcal{L}=\mathcal{L}_{\text{reg}}+\gamma\cdot\mathcal{L}_{\text{SeSA}}+\theta\cdot\mathcal{L}_{\text{CCC}},
\label{eq5}
\end{equation}
where $\mathcal{L}_{\text{reg}}$ is the base regressor loss as following:
\begin{equation}
\mathcal{L}_{\text{reg}}=\frac{1}{|\mathcal{D}^l|} \sum_{(x,y)\in\mathcal{D}^l} \big(y-\Psi_t(\Phi_t(x))\big)^2,
\label{eq6}
\end{equation}
$\gamma$ and $\theta$ are the hyperparameters to balance the loss weights.

\begin{table*}[htb!]
    \centering
    \caption{Comparsion on recalibration performance.}

    \begin{tabular}{c|c|ccc|c|cc}
    \hline CC / R$^2$ &NaiveDecoder*  &MMD* &DA-DCF* &CycleGAN &MeanTeacher* &Ours  \\
		\hline
	 C0 & 0.9776 / 0.9265 & - / - & - / - & - / - & - / - & - / - \\
	 C1 & 0.7938 / 0.6004 & 0.8165 / 0.6349 & 0.8385 / 0.6203 & 0.8315 / 0.6479  &0.8349/ 0.6609 & \textbf{0.8716 / 0.7281}  \\
	 C2 & 0.6594 / 0.4055  & 0.6711 / 0.4087 & 0.6077 / 0.4164 & 0.7012 / 0.4557  &0.8459 / 0.6630 & \textbf{0.8569 / 0.6807}  \\
	 C3 & 0.5943 / 0.3216  & 0.6209 / 0.3023 & 0.6301 / 0.3011 & 0.6578 / 0.3628  &0.8001 / 0.6261 & \textbf{0.8515 / 0.6735}  \\
    \hline
	 J0 & 0.9231 / 0.8867 & - / - & - / - & - / - & - / - & - / - \\
	 J1 & 0.8189 / 0.6641 & 0.8298 / 0.6774 & 0.8309 / 0.6762 & 0.8119 / 0.6504 & 0.8303 / 0.6926 & \textbf{0.8474 / 0.7166}  \\
	 J2 & 0.8171 / 0.6224 & 0.8147 / 0.6327 & 0.7875 / 0.5901 & 0.7974 / 0.6361  &0.8201 / 0.6491 & \textbf{0.8245 / 0.6675} \\
	 J3 & 0.7748 / 0.6095 & 0.7939 / 0.6182 & 0.7584 / 0.5855 & 0.7782 / 0.5963  &0.8142 / 0.6712 & \textbf{0.8131 / 0.6813}
  \\
    \hline
	 M0 & 0.9003 / 0.8832 & - / - & - / - & - / - & - / - & - / - \\
	 M1 & 0.7726 / 0.6482 & 0.7798 / 0.6546 & 0.7789 / 0.6561 & 0.7821 / 0.6572 & 0.7745 / 0.6484 & \textbf{0.7899 / 0.6617}  \\
	 M2 & 0.6766 / 0.4917 & 0.6834 / 0.5057 & 0.6784 / 0.5015 & 0.6794 / 0.5112 & 0.7737 / 0.5959 & \textbf{0.7880 / 0.6278}  \\
	 M3 & 0.5845 / 0.4228 & 0.4572 / 0.3328 & 0.4231 / 0.3043 & 0.4951 / 0.3679 & 0.7040 / 0.5358 & \textbf{0.7357 / 0.5624} \\
    \hline
    \end{tabular}

    \vspace{-10px}
    \label{tab:1}  
\end{table*}

\section{EXPERIMENTS}
In this section, we first introduce our base experimental settings and then conduct extensive experiments aiming to address the following three research questions (RQs):
\begin{itemize}
\item \textbf{RQ1:} Does our proposed SSAR achieve state-of-the-art recalibration performance?
\item \textbf{RQ2:} Do the three key factors we discussed in Section~\ref{sec:pre} actually improve the recalibration accuracy?
\item \textbf{RQ3:} Is our SSAR robust to various experimental settings?
\end{itemize}

\subsection{Experimental Settings}

\subsubsection{\textbf{Datasets}}
In our experiments, we adopt three nonhuman primates (NHPs) datasets published in \cite{long,cyclegan}, which named Chewie, Jango and Mihili respectively.
Chewie performed an 8-direction centre-out (CO) reaching task, Mihili performed a random-target (RT) reaching task, and Jango performed an isometric (ISO) wrist task. 
The published data were recorded by the same device across different days (with a maximum span of three months).
All tasks have three time points: 'trial start', 'go cue time', and 'trial end'.
We extracted the trials from 'go cue time' to 'trial end', and adopted multiunit threshold crossings on each channel instead of isolated single units for our experiments. 
To obtain a smoothed firing rate, we applied a Gaussian kernel (S.D.=100 ms) to the spike counts in 50 ms and normalized the neural data by z-score.
Each of our experiments is run on the cross-session data recorded on two different days of three datasets, where we regard the former-day data as the source domain and the latter-day data as the target domain.
Besides, we randomly split the data on the target domain into labeled target data and unlabeled target data for our semi-supervised domain adaptation.

\subsubsection{\textbf{Metrics}}
We evaluate the models after recalibration on the unlabeled target data.
We adopt the correlation coefficient(CC) and the coefficient of determination (R2), which are most widely used in neural decoding tasks, as our evaluation metrics.

\subsubsection{\textbf{Implementing Details}}
In the implementation of our proposed SSAR, we adopt a 3-layer MLP with relu activations as our feature extractor $\Psi$.
The dimensions of the hidden layers are 64, 32, and 16 sequentially.
The regressor of our SSAR is a simple linear layer.
We set the hyperparameters in SSAR as: $\alpha=1$, $\beta=0.1$, $\gamma=1$, $\theta=0.01$, and the number of subdomains $C=8$.
Unless otherwise noted, the proportion of the labeled target data to the total target data is set to $10\%$.
In recalibration, we adopt the Adam optimizer with \textit{lr} as $10^{-4}$, \textit{betas} as $(0.9, 0.999)$ and \textit{weight\_decay} as $5\times 10^{-4}$.
The batch size is set to 128 and the training epoch is 500.
Note that all experiments are carried out in the above setup.

\subsubsection{\textbf{Compared Methods}}
To compare the recalibration performance, we divide existing methods into three categories and choose representative methods from each category, as follows:
\begin{itemize}

\item Supervised method: a decoder which is directly trained on both the source data and the labeled target data, abbreviated as NaiveDecoder for convenience.
\item Unsupervised domain adaptation-based methods: MMD\cite{mmd1}, DA-DCF\cite{DA-DCF}, and CycleGAN\cite{cyclegan}, which utilize the source data and the unlabeled target data for recalibration.
\item Semi-supervised method: MeanTeacher\cite{meanteacher}, which performs semi-supervised learning on the source domain and target domain.
\end{itemize}

\subsection{Recalibration Performance Comparison (RQ1)}
To compare the recalibration performance of our proposed SSAR and other methods, we choose four-day data (C0 $\sim$ C3 for Chewie, J0 $\sim$ J3 for Jango, and M0 $\sim$ M3 for Mihili) for the experiment, and the results are shown in Table~\ref{tab:1}.
We regard the earliest C0, J0, and M0 as the source domains.
The lines of C0, J0 and M0 show the accuracy of the decoder trained and evaluated on current-day data, which indicates the optimum recalibration performance ideally.
The target domains include the later data with short (C1, J1, M1), medium (C2, J2, M2) and long (C3, J3, M3) time spans from the earliest days.
The corresponding lines show the recalibration performance with different recalibration methods.
Note that the results with '*' are reproduced by ourselves with the model parameters consistent with SSAR, and the results without '*' are collected from the corresponding cited papers
From Table~\ref{tab:1}, we can conclude the following three points:
\begin{enumerate}
\item The performance of NaiveDecoder degrades significantly from source domains to target domains. This observation is consistent with~\cite{kalman} that the mapping between neural signals and behaviors changes over days.
\item The existing recalibration methods can mitigate the performance degradation across days partially. However, they still have room for improvement.
\item Our proposed SSAR achieves the best performance among all existing recalibration methods on all datasets, which demonstrates its superiority.
\end{enumerate}

\subsection{Ablation Study (RQ2)}
To analyze the effects of our proposed three keys for accurate recalibration corresponding to three components in SSAR, we conduct extensive ablation experiments with C0 as the source domain and C1 as the target domain.
Specifically, the global alignment corresponds to the loss function $\mathcal{L}^*_{\text{SeSA}}$, and $\alpha$ controls its weight.
The conditional speed alignment corresponds to the loss function $\mathcal{L}^i_{\text{SeSA}}$ ($i$ ranges from $1$ to $C$), and $\beta$ controls its weight.
The feature-label consistency corresponds to the loss function $\mathcal{L}_{\text{CCC}}$, and $\theta$ controls its weight.
In order to explore their effects, we turn off the three components by setting $\alpha=0$, $\beta=0$, and $\theta=0$ respectively.
The experimental results are shown in Table~\ref{tab:2}, where G., S., and C. represent global alignment, conditional speed alignment, and feature-label consistency, respectively.
The marks $\checkmark$ and - indicate turning on or off the components.

From Table~\ref{tab:2} we can observe that all of the three components benefit the recalibration performance, proving the effectiveness of each component.
When turning on two components at the same time, the recalibration performance is better than one component alone.
Apparently, SSAR with the three components achieves the best recalibration performance.
More specifically, G. provides a basis for recalibration performance.
Additionally adopting S. can improve the performance significantly.
Further additional use of C. can enhance the performance even more.
This verifies our hypothesis in Section~\ref{sec:pre} that conditional speed alignment and the feature-label consistency between the  and the  can cover the shortage of global alignment.

\begin{table}[t]
\centering
    \caption{Ablation results of G. , S. and C.}
	\begin{tabular}{l|ccc|cc}
	\hline Method & G. & S. & C. & CC & R$^2$ \\
	\hline SSAR($\alpha=0,\beta=0,\theta=0$) & - & - & - & $0.8031$ & $0.6110$ \\
	SSAR($\alpha=0,\beta=0$) & - & - & $\checkmark$ & $0.8309$ & $0.6332$ \\
	SSAR($\alpha=0,\theta=0$) & - & $\checkmark$ & - & $0.8256$ & $0.6436$ \\ 
	SSAR($\beta=0,\theta=0$) & $\checkmark$ & - & - & $0.8165$ & $0.6349$ \\
	SSAR($\theta=0$) & $\checkmark$ & $\checkmark$ & - & $0.8439$ & $0.6943$ \\
	SSAR($\beta=0$) & $\checkmark$ & - & $\checkmark$ & $0.8462$ & $0.7019$  \\
	SSAR($\alpha=0$) & - & $\checkmark$ & $\checkmark$ & $0.8415$ & $0.6618$ \\
 	SSAR & $\checkmark$ & $\checkmark$ & $\checkmark$ & $\mathbf{0.8716}$ & $\mathbf{0.7281}$ \\
	\hline
	\end{tabular}
    \vspace{-10px}
    \label{tab:2}
\end{table}

\subsection{Robustness Analysis (RQ3)}
We explore the robustness of our proposed SSAR from two perspectives: the time span between the source and target domain and the proportion of labeled data in the target domain.

\subsubsection{Time Span}
\begin{figure}[h]
\centering{\includegraphics[width=1\linewidth]{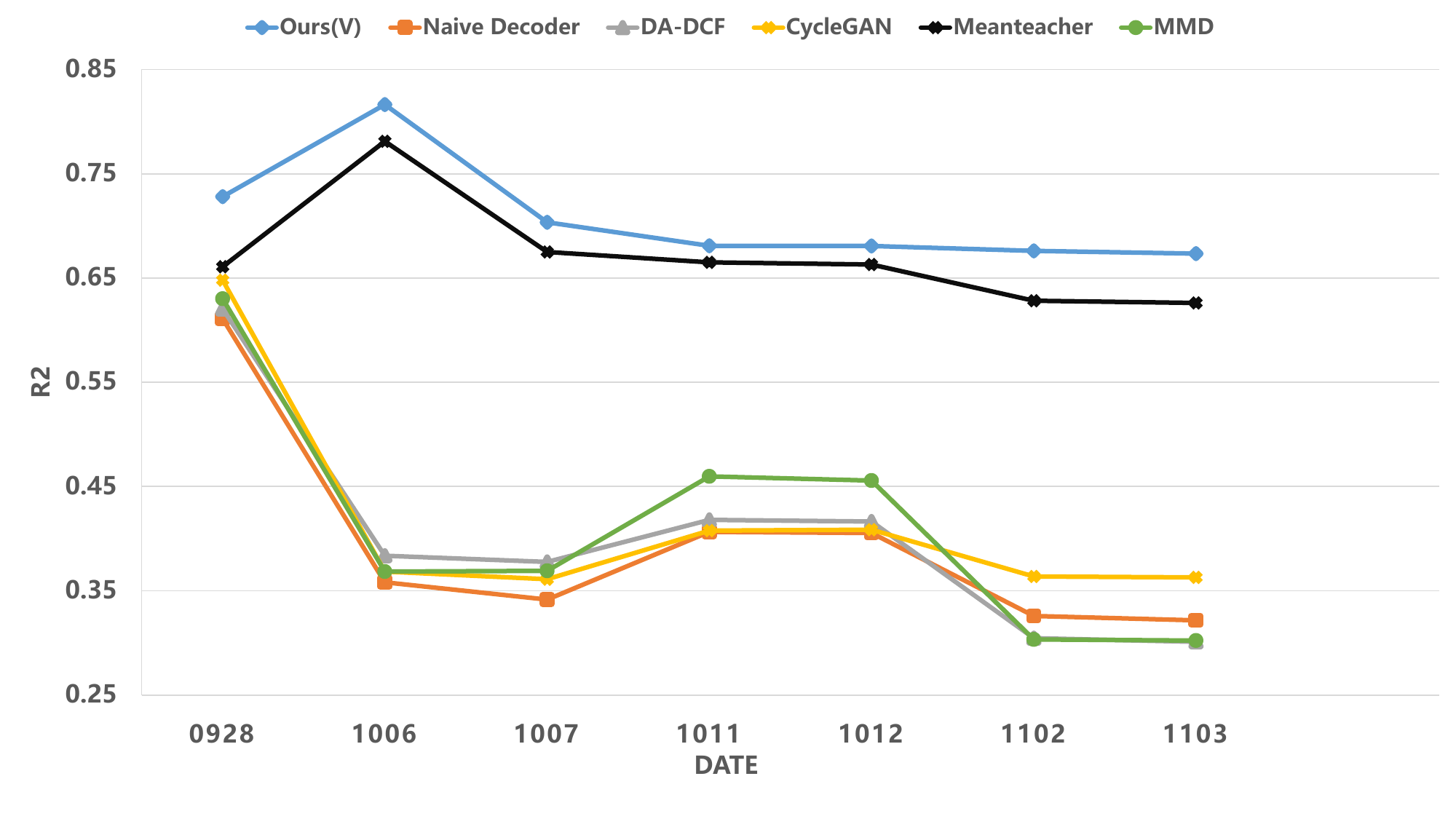}}
\vspace{-20px}
\caption{Robustness analysis on time spans}
\label{fig4}
\end{figure}

Intuitively the neural signals change progressively over time, and a long time span usually makes recalibration more difficult.
For verifying the robustness of our method to time span, we evaluate SSAR on Chewie with a maximum time span of two months, as shown in Fig.~\ref{fig4}.
In general, the performance of all recalibration methods drops with longer time spans.
However, it is worth noting that our SSAR outperforms other methods in any time span, and even on the last day our SSAR has better recalibration performance than the others on the first day.
This proves that SSAR can actually achieve the goal of long-term stable BCIs.

\subsubsection{Proportion of Labeled Data}
\begin{figure}[h]
\centering{\includegraphics[width=1\linewidth]{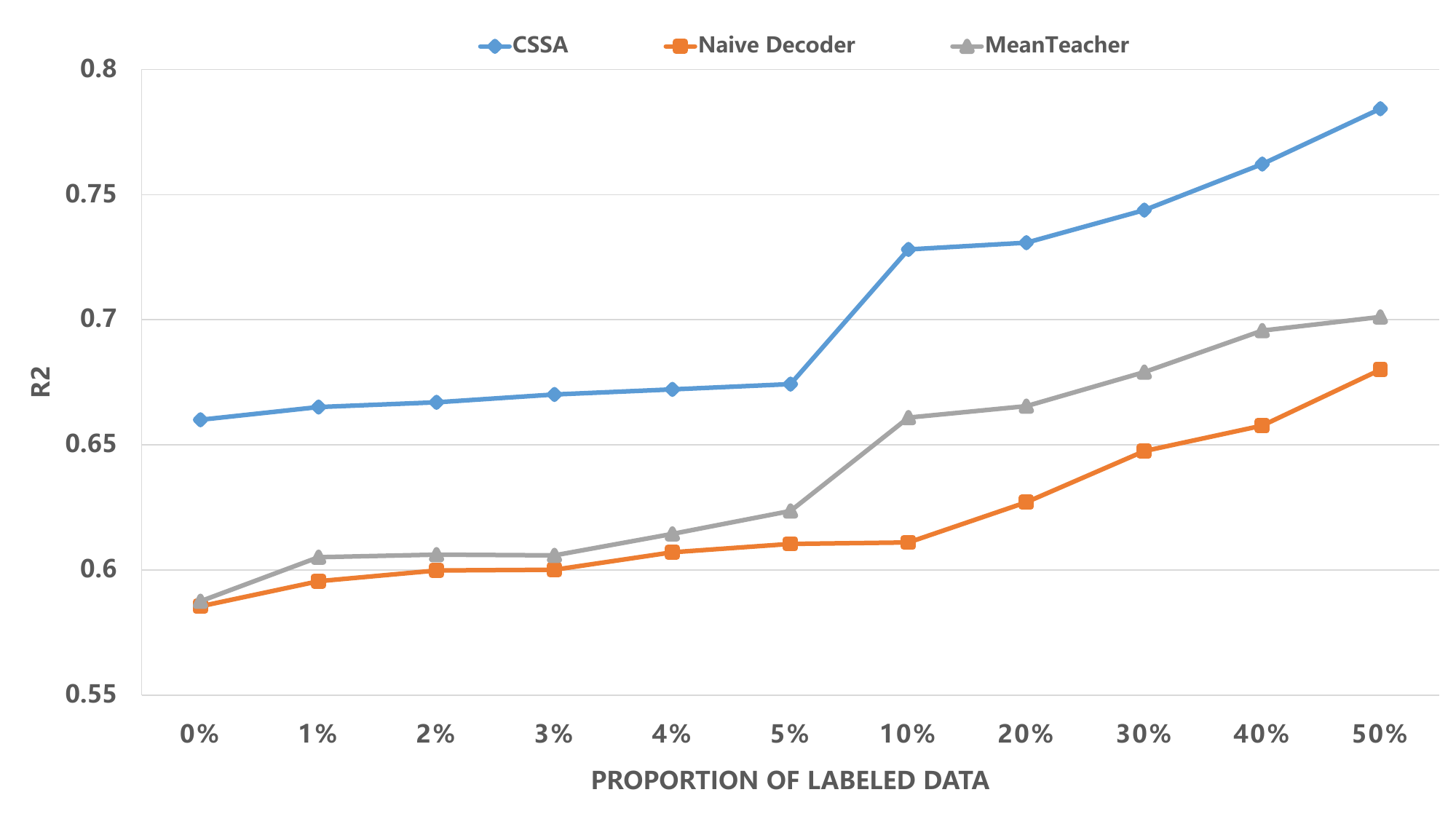}}
\vspace{-15px}
\caption{Robustness analysis on proportions of labeled data}
\label{fig5}
\end{figure}

We experiment with different proportions of labeled data in the target domain to investigate their influence on SSAR, where C0 is the source domain and C1 is the target domain.
We first take the proportion from $0\%$ to $5\%$ to explore the performance of SSAR with extremely few labeled data in the target domain.
Then we take the proportion from $10\%$ to $50\%$ to test the robustness of SSAR at a larger scale.
As shown in Fig.~\ref{fig5}, our SSAR works well with extremely small proportion of labeled data, and our SSAR remains effective even in the absence of labels.
Besides, the recalibration performance of SSAR improved with the increase of the labeled data proportion.
And with the same proportion, our SSAR always outperforms the other methods.
All of these observations indicate the generalizability and practicality of SSAR.

\section{Discussion}
With the same settings in Section~\ref{sec:pre}, we adopt PCA to visualize the features aligned by our proposed SSAR from the views of speed and direction, as shown in Fig.~\ref{fig1}.
The MMD method can be considered as a special case of SSAR($\beta=0,\theta=0$), which only performs global alignment.
Comparing the visualization of SSAR with that of MMD, we can conclude that our proposed conditional speed alignment and feature-label consistency lead to a more aligned feature distribution on the basis of global alignment, hence improving the recalibration performance effectively.

The limitation of our method is that it is not an online approach and still requires data collection before the offline recalibration.
This means that the recalibration is not a closed-loop process, and the users cannot get real-time feedback during the offline recalibration.
These limitations negatively affect the user experience.
In our future work, we aim to solve the above limitations in BCIs recalibration.

\section{CONCLUSION}
In this paper, inspired by the realistic recalibration scenarios where a small amount of the target labeled data are usually easily accessible, we propose a novel Speed-enhanced Subdomain Adaptation Regression (SSAR) for recalibration to achieve long-term stable neural decoding in BCIs.
Specifically, we first conduct pre-experiments to observe the patterns of neural signal changes across days and then analyze the shortages of global alignment in previous work.
Based on the results of the pre-experiments, we find three key factors for accurate recalibration: global alignment, conditional speed alignment and feature-label consistency.
Afterward, we propose SSAR with Speed-enhanced Subdomain Alignment (SeSA) and Contrastive Consistency Constraint (CCC).
SeSA performs global alignment and conditional speed alignment by dividing data with continuous labels into several subdomains.
CCC constrains the consistency between features and labels by contrastive learning which enhances the alignment of SeSA.
We conduct sufficient quantitative and qualitative experiments which demonstrate the superiority and the robustness of our method.
Besides, our extensive ablation experiments verify the effectiveness of each component in our method.


\end{document}